\documentstyle[sprocl]{article}

\def\be{\begin{equation}}
\def\ee{\end{equation}}
\def\bea{\begin{eqnarray}}
\def\eea{\end{eqnarray}}

\begin{document}

\title{LASER INTERFEROMETER GRAVITATIONAL WAVE DETECTORS}

\author{ A.M. SINTES }

\address{Albert-Einstein-Institut, Max Planck Institut f\"ur 
Gravitationsphysik.\\
 Schlaatzweg 1, D-14473 Potsdam. Germany}


\maketitle\abstracts{I give a brief introduction on gravitational wave laser
interferometers, possible 
detectable sources from the ground  and noise in the detectors.
}

Gravitational wave  research started in the early 1960's 
thanks to the pioneering work of Weber \cite{W}. Since that time, there
has been an ongoing research effort to develop detectors of sufficient
sensitivity to allow the detection of  waves from astrophysical sources.

The effect of a gravitational wave 
 of amplitude $h$ is to produce a strain in 
space  given by $\Delta L/L=h/2$. The magnitude of the problem facing
researchers in this area can be appreciated when one realizes that
theorists predict that for a reasonable 
\lq event rate' one should aim for a strain sensitivity of $10^{-21}$
to  $10^{-22}$. This means  that if we were monitoring the separation
of two free test masses of one meter apart, the change in separation
would be $10^{-21}$ m. Such figures should give the reader a feeling
for the enormity of the experimental challenge facing those developing
gravitational wave detectors.

The different types of detectors can be classified into two main
categories: those using laser interferometers with very long 
arms \cite{Sa} 
and those using resonant metallic masses that may be cooled to
ultra-low temperatures \cite{Gibbons71,Astone93}. The first  have the ability
to measure the gravitational wave induced strain in a broad frequency band
(expected to range from 50 Hz up to perhaps 5 kHz), while the latter measure
the gravitational wave Fourier components around the bar's resonant 
frequency (usually near 1 kHz), with a bandwidth of order a few Hz.
For resonant mass antennas, the fundamental limitation to their sensitivity
comes from the thermal motion of the atoms that can be reduced by cooling
them to temperatures of order 50 mK; existing antennas can then achieve a 
sensitivity  $10^{-19}$
to  $10^{-20}$. 

In the early 1970's emerged the idea that laser interferometers might have
a better chance of detecting gravitational waves. Detailed studies were
carried out by Forward and his group \cite{Fo} and Weiss \cite{Wei}.
Since then, several groups have developed prototype interferometric detectors
at Glasgow (10 meter Fabry-Perot), Garching (30 m delay line), MIT,
Caltech (40 m Fabry-Perot) and Tokyo. 
Projects to build long arm laser interferometers have also been funded.
The construction
of the  detectors has already been started and is rapidly progressing.
There is the  LIGO project \cite{Abra1} to build two 4 km detectors in USA, 
the VIRGO project \cite{Br} to build  a 3 km detector in Italy, the GEO-600 
project \cite{Ho} to build a 600 m detector in Germany and the
TAMA-300 project to build a 300 m  detector in Japan. 
Not yet funded is the Australian project of a 500 m  detector to  be
built near Perth where the site would allow a later extension to a 
3 km arms.
 GEO-600, LIGO, TAMA-300 and VIRGO are scheduled to be 
 completed by the end of this
 century. Observations may begin in 2000 or 2001 and advanced detector 
 sensitivity will be reached sometime between 2005 and 2010.
 
 There appear to be several advantages to these types of detectors.
Today, prototype interferometers \cite{Abra,Ro} are routinely  operating at a 
displacement  noise level  of a few
 times $10^{-19}$ m$/\sqrt{{\rm Hz}}$
over a frequency range from 200 Hz to 1000 Hz corresponding to an 
rms gravitational-wave amplitude noise level of 
$h_{rms}\sim 2\times 10^{-19}$.

Laser interferometer detectors have the potential to see the full 
range of gravitational wave 
signals: periodic, burst, quasiperiodic and stochastic. 
There are different possible  sources 
detectable from the ground. Strong bursts from supernova and
gravitational collapse are amongst the most powerful sources
 of gravitational radiation. Coalescing binaries of compact objects
consisting either of neutron stars or black holes 
will be a strong source of gravitational waves at high-frequency
during the last few minutes before the stars coalesce. They will
emit an almost monochromatic \lq chirp' signal whose frequency
increases with time in a predictable way. Pulsars (individual neutron
stars) can radiate  signals
 that we expect to be largely monochromatic with their 
frequency being modulated by Doppler shifts due to the relative
motions between emitter and receiver.
And,  analogous to the electromagnetic microwave background
radiation, 
we have the stochastic cosmological background, which will be the result
of the cooperative effect of a variety of different sources 
(binary systems, black holes, phase-transitions, cosmic strings, etc)
superimposed on one another in a random way providing  information from
the Universe close to the Big Bang.

In order to detect these signals, we need to build extremely
quiet systems (with a very low level of noise) and, 
at the same time, we need to design an adequate data
analysis strategy.

There are different noise sources  that limit the sensitivity of the
laser-interferometer detectors. As a result of noise in the detectors,
we will be able to detect the gravitational wave signals only with a certain
probability and estimate parameters of the signals only with a certain
confidence.

The noise characterization within the interferometer is of two types:
non-Gaussian noise and Gaussian noise. The first one may occur
several times per day from events such as strain release in the
suspension systems. The only way  to remove these non-Gaussian 
events is through coincidence comparison of each of the signals
from different interferometers. Gaussian noise is unlikely to generate
noise bursts and can be characterized by an amplitude spectral 
density.

The performance of any gravitational wave detector  is intimately related
to the noise sources, in particular  with the Gaussian noise.
The dominant noise sources have been studied using prototype 
interferometers.

The stochastic noise can be modeled as a sum of several 
contributions \cite{Kro}.
At frequencies above 200 Hz, the noise in the interferometer is
dominated by the shot noise which is  due to the statistical fluctuations
in the number of photons detected at the output.
At frequencies between 70 Hz and 200 Hz, 
thermal noise from three sources dominate:
the vibration of the test masses (\lq mirror modes'),  
the vibration of the suspension wires (\lq violin modes'),
and the low frequency oscillations of the
pendulum suspensions (\lq pendulum modes'). Notice that the resonance vibrations
of the mirrors and the suspensions do appear at higher frequencies
when the shot noise dominates. At frequencies lower than 70 Hz,
the noise is dominated by the seismic noise originated  from the
ambient vibration of the ground due to the geological activity of the
Earth, wind forces coupled to trees and buildings, and man-made sources.
Also, there is 
 quantum noise (which follows from
the indeterminacy of the position of the test masses due to the
Heisenberg uncertain principle) and  seismic gravity  gradient noise
(also called \lq Newtonian noise').
The above noise sources may be considered among the most important,  but
other sources  of noise cannot be ignored. 

The characterization of the noise spectral density in the
detectors is very important for  data analysis. 
Its accurate knowledge is
 transcendental  for the detection of 
signals.
In the measured noise spectrum of the different interferometer prototypes,
in addition to  stochastic noise, we observe
 peaks due to external interference, where the amplitudes are not stochastic.
We have already designed an algorithm  able to get rid of them \cite{AS2},
with the purpose to remove as many interference features as possible, so
that the interferometer sensitivity is limited by the
genuinely stochastic noise. 
   
 For further reading on this topic, we refer the reader to \cite{S,T}.

\section*{Acknowledgments}
This work was partially supported by the European Union, 
 TMR Contract
No. ERBFMBICT972771.


\section*{References}
\begin{thebibliography}{99}

\bibitem{W} Weber J., {\em Phys. Rev.} {\bf 117}, 306 (1960).
\bibitem{Sa} Saulson P.R., {\em Fundamentals of interferometric 
gravitational wave detectors}, (World Scientific, Singapore, 1994)

\bibitem{Gibbons71}  Gibbons G.W.,  Hawking S.W.,  {\em Phys. Rev.} D 
{\bf 4}, 2191 (1971)
\bibitem{Astone93}  Astone P., {\em et al.},  {\em Phys. Rev.}  D  {\bf 47},
 362 (1993).
\bibitem{Fo} Forward R.L.,  {\em Phys. Rev.} D {\bf 17}, 379
(1978)
\bibitem{Wei} Weiss R.,  {\em MIT, Quart Progress Report} n. 105 (1972)

\bibitem{Abra1} Abramovici A., Althouse W.E., Drever R.W.P., 
Gursel Y., Kawamura S.,
 Raab F.J.,  Shoemaker D., Sievers L.,
Spero R.E., Thorne K.S., Vogt R.E., Weiss R., Whitcomb S.E., Zucker M.,
{\em Science} {\bf 256}, 325 (1992)

\bibitem{Br} Bradaschia C., Del Fabbro R., 
 Di Virgilio A., Solimeno S.,
Giazotto A., Kautzky H., Montelatici V., Passuello D., Brillet A.,  
Cregut O., Hello P., Man C.N., 
Manh P.T., Marraud A,
 Shoemaker D., Vinet J.Y., Barone F., Di Fiore L., Milano L.,
 Russo G.,  Aguirregabiria J.R., Bel H., Duruisseau J.P.,
 Le Denmat G., Tourrenc Ph., Capozzi M., Longo M., Lops M.,
 Pinto I., Rotoli G., Damour T., Bonazzola S., Marck J.A.,
 Gourghoulon Y., Holloway L.E., Fuligni F., Iafolla V., Natale G.,
 {\em Nucl. Instr. Meth. Phys. Res.}, A {\bf 289}, 518 (1990)


\bibitem{Ho} Hough J., {\em et al.},
{\em GEO-600: Proposal for a
600 m Laser-Interferometric  Gravitational Wave Antenna} in
{\em Proceedings of the 7th Marcel Grossman Meeting}, (1994)

\bibitem{Abra} Abramovici A., Althouse W., Camp J., Durance D.,
Giaime J.A., Gillespie A., Kawamura S., Kuhnert A.,
Lyons T., Raab F.J., Savage Jr. R.L., Shoemaker D., Sievers L.,
Spero R., Vogt R., Weiss R., Whitcomb S., Zucker M.,
{\em Physics Letters} A {\bf 218}, 157 (1996)

\bibitem{Ro} Robertson D.I., Morrison E., Hough J., Killbourn S.,
Meers B.J.,  Newton G.P.,  Robertson N.A.,  Strain K.A.,  Ward H., 
{\em  Rev. Sci. Instr.} {\bf 66}, 4447 (1995)

\bibitem{Kro} Kr\'olak A.,  {\em  Acta Cosmologica}  22-1 (1996)

\bibitem{AS2} Sintes A.M.,  Schutz B.F.,  
{\em Phys. Rev.} D (Dec. 1998).  gr-qc/9810004.

\bibitem{S} Schutz B.F.,   in 
{\it Relativistic Gravitation and Gravitational Radiation},
eds. Marck J.A. \& Lasota J.P., (Cambridge
University Press,  1997)

\bibitem{T} Thorne K.S.,   in
{\it Proceedings of Snowmass 1994 Summer Study on Particle and Nuclear 
Astrophysics and Cosmology},
eds. Kolb E.W. \& Peccei R., (World Scientific, Singapore, 1995) 


\end{thebibliography}
\end{document}